\documentclass[a4paper,11pt]{article}
\usepackage{jheppub} % for details on the use of the package, please see the JINST-author-manual
\usepackage{lineno}
\usepackage{braket}
\newcommand{\dis}[1]{\begin{equation}\begin{split}#1\end{split}\end{equation}}

\title{\boldmath Probing the mixing between sterile and tau neutrinos in the SHiP experiment}

\author[a]{Ki-Young Choi,}
\author[b]{Sung Hyun Kim,}
\author[c]{Yeong Gyun Kim,}
\author[b]{Kang Young Lee,}
\author[d]{Kyong Sei Lee,}
\author[b]{Byung Do Park,}
\author[b]{Jong Yoon Sohn,}
\author[a]{Seong Moon Yoo}
\author[b]{and Chun Sil Yoon}

\affiliation[a]{Department of Physics and Institute of Basic Science, Sungkyunkwan University, Suwon 16419, Korea}
\affiliation[b]{Department of Physics Education \& Research Institute of Natural Science,
Gyeongsang National University, Jinju 52828, Korea}
\affiliation[c]{Department of Science Education, Gwangju National University of Education, Gwangju 61204, Korea}
\affiliation[d]{Center for Extreme Nuclear Matters, Korea University, Seoul 02841, Korea}

\emailAdd{kiyoungchoi@skku.edu}
\emailAdd{castledoor@skku.edu}

\abstract{We study the expected sensitivity to the mixing between sterile and tau neutrinos directly from the tau neutrino disappearance in the high-energy fixed target experiment. Here, the beam energy is large enough to produce tau neutrinos at the target with large luminosity. During their propagation to the detector, the tau neutrino may oscillate into sterile neutrino. By examining the energy spectrum of the observed tau neutrino events, we can probe the mixing between sterile and tau neutrinos directly. In this paper, we consider Scattering and Neutrino Detector (SND) at SHiP experiment as a showcase, which uses 400 GeV protons from SPS at CERN, and expect to observe 6,300 tau and anti-tau neutrinos from the $2\times 10^{20}$ POT for 5 years operation. Assuming the uncertainty of 10\%, we find the sensitivity $|U_{\tau 4}|^2 \sim 0.08$\, (90\% CL) for $\Delta m_{41}^2 \sim 500\ \mathrm{eV}^2$ with 10\% signal-to-background ratio. We also consider a far SND at the end of the SHiP hidden sector detector, in which case the sensitivity would be enhanced to $|U_{\tau 4}|^2 \sim 0.02$.
}

\begin{document}
\maketitle
\flushbottom

%%%%%%%%%%%%%%%%%%%%%%%%%%%%%%%%%%
\section{Introduction}
\label{Intro}
%%%%%%%%%%%%%%%%%%%%%%%%%%%%%%%%%%

One of the simplest extensions to explain the neutrino mass is the introduction of additional right-handed neutrinos in the standard model~\cite{Minkowski:1977sc,Ramond:1979py,Gell-Mann:1979vob,Yanagida:1979as,Mohapatra:1979ia}.
These additional neutrinos should be free from the weak interaction~\cite{Janot:2019oyi}, and thus they are called \textit{sterile}. The sterile neutrinos may also explain the anomalies in the short baseline neutrino oscillation experiments~\cite{LSND:2001aii,MiniBooNE:2020pnu}, and solve the cosmological problems of dark matter and baryon asymmetry~\cite{Dasgupta:2021ies}.

The sterile neutrinos do not have weak interaction, however, they can communicate with visible sector through the mixing with active neutrinos at low energy. The resulting oscillation has been the main subject to search and constrain the sterile neutrinos~\cite{Dentler:2018sju,Acero:2022wqg}.
The mixing of sterile neutrino with electron and muon neutrino is constrained with a bound of $|U_{e4}|^2\lesssim 0.04$ (90\% CL) for  $\Delta m^2_{41}\sim 10^{-2}\ev^2$ and $|U_{\mu4}|^2\lesssim 10^{-2}$ for  $10^{-2}\ev^2\lesssim\Delta m^2_{41}\lesssim 1\ev^2$~\cite{MINOS:2020iqj}.

About the mixing between sterile and tau neutrinos, two-neutrino oscillations of tau neutrinos are constrained by tau neutrino appearance. For instance, the constraint of $\nu_\mu\rightarrow \nu_\tau$ appearance from the NOMAD~\cite{NOMAD:2001xxt} and CHORUS experiments~\cite{CHORUS:2007wlo} give $4|U_{\mu 4}|^2|U_{\tau 4}|^2 < 5\times 10^{-4}$ for $\Delta m^2_{41}\gtrsim 100\ev^2$ while the OPERA experiment~\cite{OPERA:2015zci}, with its long baseline, probe it down to the mass $\Delta m^2_{41}\gtrsim 0.02\ev^2$ with a bound $4|U_{\mu 4}|^2|U_{\tau 4}|^2<0.116$.
Constraints on $|U_{\tau 4}|^2$ only, in contrast, are much weaker than other active flavors since the best way to constrain the mixing to date is through the indirect disappearance of muon neutrinos under a long baseline~\cite{Acero:2022wqg,MammenAbraham:2022xoc}. For instance, atmospheric neutrino experiments can provide constraints as $|U_{\tau 4}|^2<0.18$ in Super-Kamiokande~\cite{Super-Kamiokande:2014ndf} and $|U_{\tau 4}|^2<0.15$ in IceCube-DeepCore~\cite{IceCube:2017ivd}, for $\Delta m^2_{41}\gtrsim 1\ev^2$. 
Also, this mixing can be probed directly in neutral current interactions at the NOvA~\cite{NOvA:2021smv} and DUNE experiments~\cite{Coloma:2017ptb}, but sensitivities are weaker than the sensitivity from IceCube-DeepCore.

In this paper, to overcome such insensitive nature of $|U_{\tau 4}|^2$, we propose a new approach to constrain the mixing between sterile and tau neutrinos focusing on the SHiP experiment. SHiP experiment is a facility that has been proposed at the SPS beam dump to search for hidden sectors beyond the SM~\cite{SHiP:2015vad,Alekhin:2015byh,Aberle:2839677,Ahdida:2654870, Albanese:2878604}. Considering SND at SHiP as a short baseline, we can search for the evidence of tau neutrino disappearance during the propagation. Thanks to the high potential to detect $\mathcal{O}(10^4)$ tau neutrino Charged-Current(CC) events, it is possible to probe the mixing between sterile and tau neutrinos directly by measuring the tau neutrino energy spectrum.

The rest of this paper is organised as follows. In Sec.~\ref{Neutrino Oscillation}, we review the neutrino oscillation with sterile neutrino, and in Sec.~\ref{spectrum} we summarise the SHiP experiment and show the expected energy spectrum of tau neutrino events.
In Sec.~\ref{sensitivity}, we describe the statistical method we employed and presents the expected sensitivity to the mixing between sterile neutrinos and tau neutrinos. In Sec.~\ref{sec:FSND}, we propose a new SND with long baseline that is placed after Hidden Sector Decay Spectrometer and estimate the enhancement of sensitivities due to the new SND. Lastly, we make a conclusion in Sec.~\ref{con}.

%%%%%%%%%%%%%%%%%%%%%%%%%%%%%%%%%%%%%%%%%%%%%%%%%%%%
\section{Neutrino Oscillation with Sterile Neutrino}
\label{Neutrino Oscillation}
%%%%%%%%%%%%%%%%%%%%%%%%%%%%%%%%%%%%%%%%%%%%%%%%%%%%
The weak eigenstates of neutrino $|\nu_\alpha\rangle$ is the linear combination of the mass eigenstates $|\nu_i\rangle$ determined by the mixing matrix $\mathbf{U}$,
\begin{equation}
    \ket{\nu_\alpha}=\sum_{i} U^*_{\alpha i}\ket{\nu_i}.
\end{equation}
In `3+1' model with single sterile neutrino as well as 3 active neutrinos in SM, the  matrix becomes $4\times 4$ matrix, which can be written as
\begin{equation}
    \mathbf{U}=\begin{pmatrix}
        U_{e 1} & U_{e 2} & U_{e 3} & U_{e 4}\\
        U_{\mu 1} & U_{\mu 2} & U_{\mu 3} & U_{\mu 4}\\
        U_{\tau 1} & U_{\tau 2} & U_{\tau 3} & U_{\tau 4}\\
        U_{s 1} & U_{s 2} & U_{s 3} & U_{s 4}
    \end{pmatrix}.
\end{equation}
This mixing matrix can be parameterized with six rotation angles and three CP phases. The rotation angles include three additional mixing angles between the sterile neutrino and each active neutrino, $\theta_{14}$, $\theta_{24}$ and $\theta_{34}$ as well as the usual three mixing angles between active neutrinos $\theta_{12}$, $\theta_{23}$ and $\theta_{13}$.

After travelling a distance $L$, the state of the  neutrino evolves as
\begin{equation}
    \ket{\nu_\alpha(t)}=\sum_{i} U^*_{\alpha i}\ket{\nu_i(t)},
\end{equation}
and the probability for the transition is given by
\dis{
P_{\alpha\beta} = |\langle\nu_\beta | \nu_\alpha(t) \rangle|^2 = \left| \sum_{i,j} U^*_{\alpha i} U_{\beta j} \langle \nu_j|\nu_i(t)\rangle\right|^2.
}
Especially, for the short baseline experiment where the mixing between active neutrinos can be ignored, the probability of the relativistic neutrino can be approximated as
\begin{equation}\label{eq:short}
    P_{\alpha \beta} = \left|\delta_{\alpha \beta} -\left(1-e^{-i\frac{\Delta m^2_{41} L}{2 E_\nu}}\right)U_{\alpha 4} U_{\beta 4}^*\right|^2,
\end{equation}
where $\Delta m_{41}^2\equiv m_4^2-m_1^2$ is the mass squared difference between the sterile neutrino and the lightest neutrino, $L$ is the baseline distance, and $E_\nu$ is the energy of the neutrino.

For tau neutrino disappearance experiments, the survival probability becomes
\begin{equation}
    P_{\tau\tau} = 1-4|U_{\tau 4}|^2 (1-|U_{\tau 4}|^2) \sin^2\left(\frac{\Delta m^2_{41} L}{4 E_\nu}\right),
\end{equation}
where $|U_{\tau 4}|^2= \sin\theta^2_{34}$, ignoring other mixing angles.

\begin{figure}
    \centering
    \includegraphics[width = 0.7\textwidth]{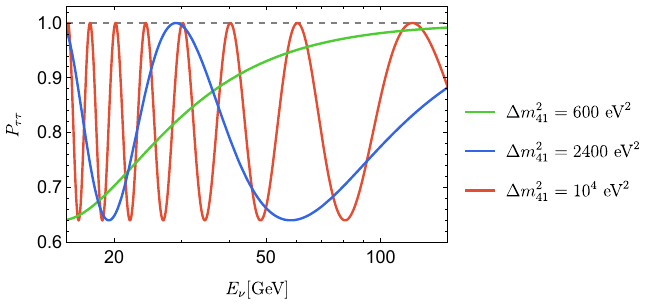}
    \caption{The survival probability of tau neutrinos, $P_{\tau\tau}$, for a mixing parameter of $|U_{\tau 4}|^2 = 0.1$ with a neutrino detector at a distance $L=30\ \mathrm{m}$ from the target. The different colors show the probability for different sterile neutrino masses:  green, blue, and red line for $\Delta m_{41}^2 = 600, 2400, 10^4\ \mathrm{eV}^2$, respectively.  The dashed gray line represents $P_{\tau\tau}=1$, for the case without mixing between tau neutrino and sterile neutrino.}
    \label{fig:oscNSND}
\end{figure}
In Fig.~\ref{fig:oscNSND}, we show the survival probability of tau neutrinos, $P_{\tau\tau}$, for a mixing parameter of $|U_{\tau 4}|^2 = 0.1$ with a neutrino detector at a distance $L=30\ \mathrm{m}$ from the target. The different colors show the probability for different sterile neutrino masses:  green, blue, and red line for $\Delta m_{41}^2 = 600,\ 2400,$ and $10^4\ \mathrm{eV}^2$, respectively. The dashed gray line represents $P_{\tau\tau}=1$, for the case without mixing between tau neutrinos and sterile neutrinos.

%%%%%%%%%%%%%%%%%%%%%%%%%%%%%%%%%%%%%%%%%%%%%%%%%%%%
\section{Tau Neutrino Spectrum at the SHiP Experiment}
\label{spectrum}
%%%%%%%%%%%%%%%%%%%%%%%%%%%%%%%%%%%%%%%%%%%%%%%%%%%%
In the SHiP experiment, the high energy protons of 400 GeV from SPS can generate all three flavors of neutrino, when the protons are stopped at tungsten target. The neutrinos propagate through the shielding to the SND located 30 m away from the proton target. Three different flavors of neutrino can be observed using CC interaction at the emulsion detector in the SND. 

While the electron and muon neutrinos are primarily produced from kaon and pion decays at the target, the tau neutrinos are mainly from the decay of the charm meson, $D_s\rightarrow \tau \nu_\tau$, and subsequent tau decays. The subsequent interaction of secondary particles, which we call cascade interaction, also can produce tau neutrinos~\footnote{Including cascade interaction, the total number of charm meson is increased by a factor of 2.3~\cite{CERN-SHiP-NOTE-2015-009}, but the detected number of $\nu_\tau$ CC events would not increase much as the number of charm meson, since the angle and the energy of secondary tau neutrino is more dispersed than the primary tau neutrino. In this paper, we use the enhancement factor of $1.3$, as suggested by Ref.~\cite{Iuliano:2776128}.}.
%\cblue{sterile neutrinos are also produced?} 
The number of tau neutrinos produced at the target can be estimated using Monte Carlo(MC) simulation of the experiment, and it could be around $10^{16}$
 for 5 years operation~\cite{Albanese:2878604}.
 %the number of tau neutrinos per unit energy $\phi_{\nu_\tau}$ and $\phi_{\bar{\nu}_\tau}$ are ..

The differential number of events of tau neutrinos with CC interaction at SND, with respect to the traveling distance and the energy of tau neutrino can be obtained by
\begin{equation}
\label{eq:Ncc0}
    \frac{d^2N}{dE_\nu dl} = n_\text{W} \frac{L_\text{W}}{L_\text{SND}} \epsilon_\mathrm{eff} \left(\phi_{\nu_\tau} \sigma_{\nu_\tau A} + \phi_{\Bar\nu_\tau} \sigma_{\Bar\nu_\tau A}\right),
\end{equation}
where $n_\text{W}$ is the number density of tungsten atom in tungsten layers with total thickness $L_\mathrm{W}$ in SND, which length is $L_\text{SND}$. $\phi_{\nu_\tau}(\phi_{\bar{\nu}_\tau})$ is the expected number of tau (anti) neutrinos passed through SND per $E_{\nu}$, $\sigma_{\nu A}(\sigma_{\bar\nu A})$ is the cross-section of (anti) neutrino-nucleus interaction, and $\epsilon_\mathrm{eff}$ is the detection efficiency of tau neutrino.
Here, the survival probability $P_{\tau\tau}(E_\nu,l) $ depends on the neutrino energy and the distance from the proton target.

%19.30 * 100 / 66.7 * 272 * 1.3 * 0.62

Instead of full MC simulation, we adopt the event rate of tau neutrino CC interaction on 3 flavor model from Ref.~\cite{Bai:2018xum}, which is the most conservative result to the best of our knowledge.  From here, we can write $\frac{dN}{dE_\nu}$ with sterile neutrino oscillation as
\begin{equation}
\label{eq:Ncc}
    \frac{dN}{dE_\nu} = \int_{L_0}^{L_0+L_\mathrm{SND}} P_{\tau\tau}(E_\nu,l) \frac{d^2N}{dE_\nu dl} dl,
\end{equation}
where $L_0$ is the distance between the target and SND.
Since the design of SND on Ref.~\cite{Bai:2018xum} is similar with the design on the technical proposal of SHiP experiment at 2015~\cite{SHiP:2015vad}, we adopt $\epsilon_\mathrm{eff}$ from the technical proposal at 2015, while $L_\text{W}=100\ \mathrm{cm}$, $L_\text{SND}=3\ \mathrm{m}$ and $L_0=30\ \mathrm{m}$, following recent designs of SND at SHiP experiment~\cite{Albanese:2878604, Ahdida:2654870}. Therefore, after 5 years operation with $2\times 10^{20}$ POT, about 6,300 tau neutrino events are expected to be observed in the 3 flavor model.

To consider the energy response of SND in the reconstructed neutrino energy $E_\mathrm{rec}$ from the true energy $E_\mathrm{\nu}$, we used the reconstructed energy spectrum as 
\begin{equation}
    \frac{dN}{dE_\text{rec}}=\int_{E_\nu =0}^{\infty} f(E_\text{rec},E_\nu ) \frac{dN}{dE_\nu } dE_\nu,
    \label{dNdErec}
\end{equation}
where the Gaussian response function $f(E_\mathrm{rec},E_\nu)$ is given by
\begin{equation}
f(E_\text{rec},E_\nu ) \equiv \frac{1}{\sqrt{2\pi} \sigma}e^{-\frac{(E_\nu - E_\text{rec})^2}{2\sigma^2}}.
\end{equation}
The energy classification from the hadron contribution determines $\sigma$, which we used as $20\%$ of $E_\nu$, \textit{i.e.} $\sigma=0.2 E_\nu$, following the result from the technical proposal of SND@LHC.~\cite{Ahdida:2750060}

Note that, apart from the energy response, the baseline uncertainty also exists. This uncertainty comes from the varying interaction lengths between the target and the protons from SPS. However, this uncertainty is relatively minor compared to the length of SND, and can be ignored in our analysis.

\begin{figure}
    \centering
    \includegraphics[width = 0.7\textwidth]{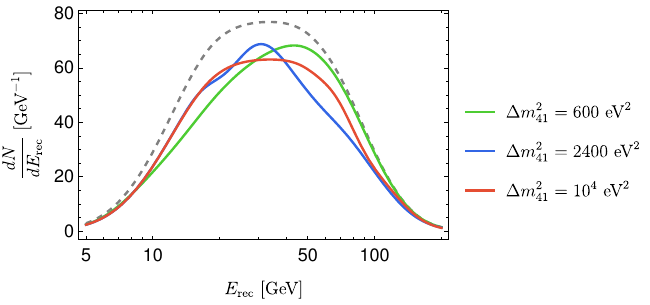}
    \caption{Differential number of distinguished $\nu_\tau$ CC events on SND with respect to the reconstructed energy $E_\mathrm{rec}$, if a mixing parameter $|U_{\tau 4}|^2 = 0.1$. Different scenarios are depicted: the green line for $\Delta m_{41}^2 = 600\ \mathrm{eV}^2$, the blue line for $\Delta m_{41}^2 = 2400\ \mathrm{eV}^2$, and the dashed red line for $\Delta m_{41}^2 = 10^4\ \mathrm{eV}^2$. The dashed gray line represents the case with $P_{\tau\tau}=1$, indicating no mixing between tau neutrino and sterile neutrino.}
    \label{fig:fluxNSND}
\end{figure}

In Fig.~\ref{fig:fluxNSND}, we  show the  number of tau neutrino + tau anti neutrino events per energy at SND with respect to $E_\mathrm{rec}$, considering the sterile neutrino oscillation for three mass squared differences,
\textit{i.e.} $\Delta m^2_{41}=600,\ 2400,$ and $10^4\,\ev^2$
with green, blue, and red colors, respectively.
We can see the depletion of the probability depending on the energy of the neutrino, due to the oscillation into sterile neutrino.

%%%%%%%%%%%%%%%%%%%%%%%%%%%%%%%%%%%%%%%%%%%%%%%%%%%%
\section{Sensitivity to the Mixing between Sterile and Tau Neutrinos}
\label{sensitivity}
%%%%%%%%%%%%%%%%%%%%%%%%%%%%%%%%%%%%%%%%%%%%%%%%%%%%
%\section{Statistical Analysis and Expected Sensitivity}\label{sec:statsen}

In this section, we show the expected sensitivity to `3+1' model using the tau neutrino disappearance at SHiP experiment. To find the expected sensitivity, we assumed that the result from SHiP experiment is in agreement with the standard three neutrino model and found the Confidence Level (CL) of the virtual result at each point of `3+1' model parameters. To find the CL, we used two different methods: 1) Wilks' theorem~\cite{10.1214/aoms/1177732360}, and 2) the profiled Feldman-Cousins (FC) method~\cite{Feldman:1997qc,NOvA:2022wnj}.

We use the well-known statistic $\Delta \chi^2$ which quantifies the fitness of the data with the corresponding model, defined as
\begin{equation}
    \Delta \chi^2 \equiv \chi^2_{\boldsymbol{\lambda}} - \chi^2_{\boldsymbol{\theta},\boldsymbol{\lambda}},
\end{equation}
where $\chi^2_{\boldsymbol a} \equiv -2 \max_{\boldsymbol a}\log L$, which maximises the likelihood function $L$ for any set of parameters $\boldsymbol a$. For a likelihood function from SHiP experiment, we use Poisson probability distributions $\mathrm{Pois}(O_i|\mu_i)$, and a likelihood function for the nuisance parameters from previous experiments $L(\boldsymbol\lambda)$, which confines their possible range with uncertainties of SHiP experiment. The combined likelihood function of a data $\boldsymbol O$ for model parameters $\boldsymbol\theta$ and nuisance parameters $\boldsymbol\lambda$ is written as
\begin{equation}
    L({\boldsymbol\theta,\boldsymbol\lambda}|\boldsymbol O) = L(\boldsymbol\lambda)\times \prod_{i \in \textbf{bin}} \mathrm{Pois}(O_i|\mu_i),
\end{equation}
where $O_i$ and $\mu_i$ are the observed and expected number of $\nu_\tau$ CC events at a $i$-th bin. The number of events at $i$-th bin $\mu_i\equiv s_i + b_i$ is the sum of signal $s_i$ and background $b_i$, which is evaluated for given set of parameters $\boldsymbol\theta$ and $\boldsymbol\lambda$.

The expected number of signal at $i$-th bin is calculated as
\begin{equation}
    s_i(\boldsymbol\theta) = (1+A)(1+\alpha_i)\int_{E_{i, \mathrm{low}}}^{E_{i, \mathrm{high}}} \left.\frac{dN}{dE_\mathrm{rec}}\right|_{\boldsymbol\theta}dE_\text{rec},
\end{equation}
where the nuisance parameters $A, \alpha_i\in \boldsymbol\lambda$ represent overall and shape uncertainties for signal, which are assumed to have a mean value of zero, while $E_{i, \mathrm{low}}$ and $E_{i, \mathrm{high}}$ are the lower and upper energy limit of $i$-th energy bin. Overall and shape uncertainties reflect the uncertainties from parton distributions, factorization and renormalization scale factor, intrinsic transverse momentum~\cite{Nelson:2012bc,Alekhin:2015byh,Bai:2018xum}, and other possible experimental uncertainties. In this paper, we assume the likelihood function of nuisance parameters as similar as a normal distribution. For simplicity, all nuisance parameters are independent of each other and share the same variance $\sigma_\mathrm{norm}^2$ during the analysis, even though the likelihood function may include non-zero covariance terms between nuisance parameters. Therefore, we rewrite $L(\boldsymbol \lambda)$ as
\begin{equation}\label{eq:nuisance_likelihood}
    L(\boldsymbol\lambda) \sim \prod_{\lambda\in \boldsymbol\lambda} \mathrm{Exp}\left[-\frac{\lambda^2}{2\sigma_\mathrm{norm}^2}\right].
\end{equation}
For the variance of nuisance parameters, we choose $\sigma_\mathrm{norm} = 10\%$ or $\sigma_\mathrm{norm} = 20\%$ in later figures. These uncertainties can be refined with the results from the DsTau Project in the future~\cite{DsTau:2019wjb}.

\begin{figure}
    \centering
    \begin{center}
        \includegraphics[width = .49\textwidth]{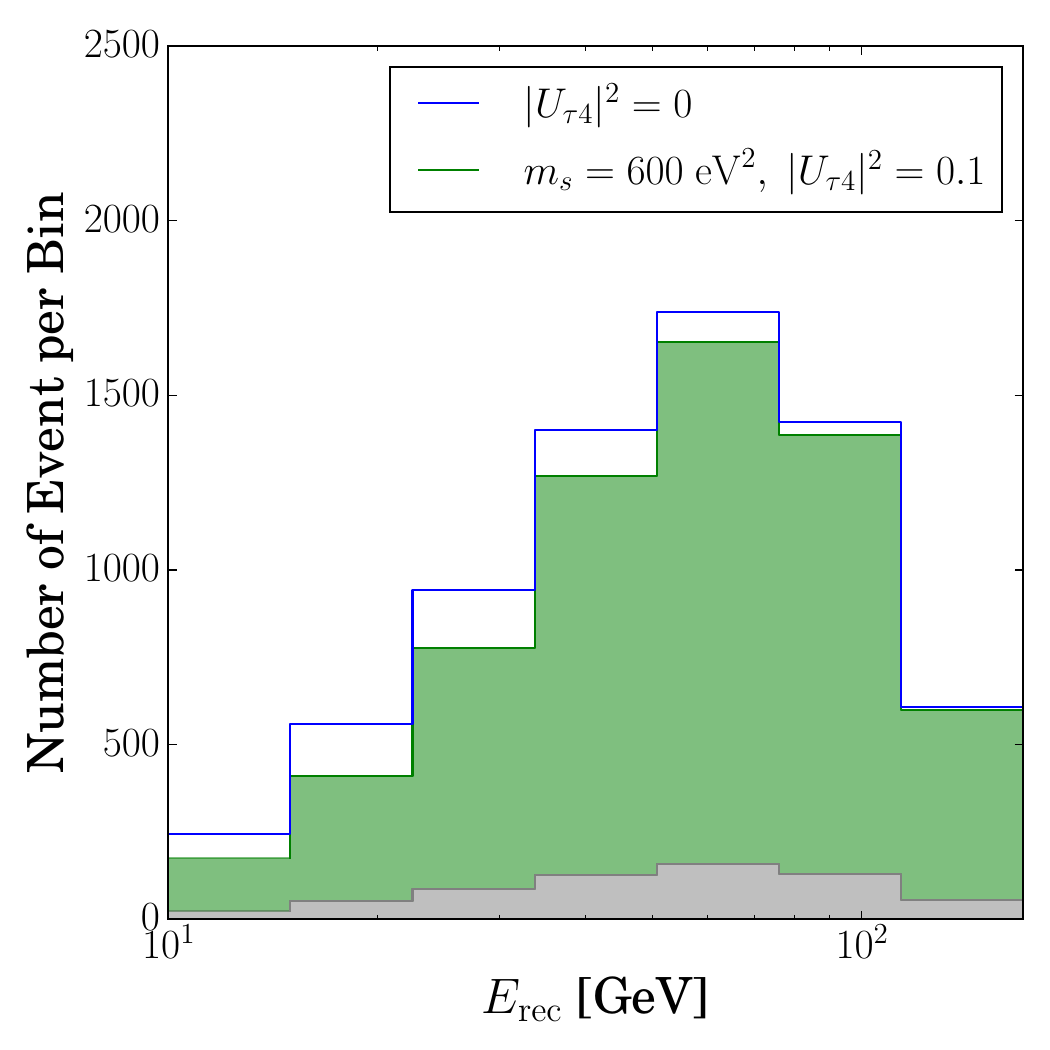}
        \includegraphics[width = .49\textwidth]{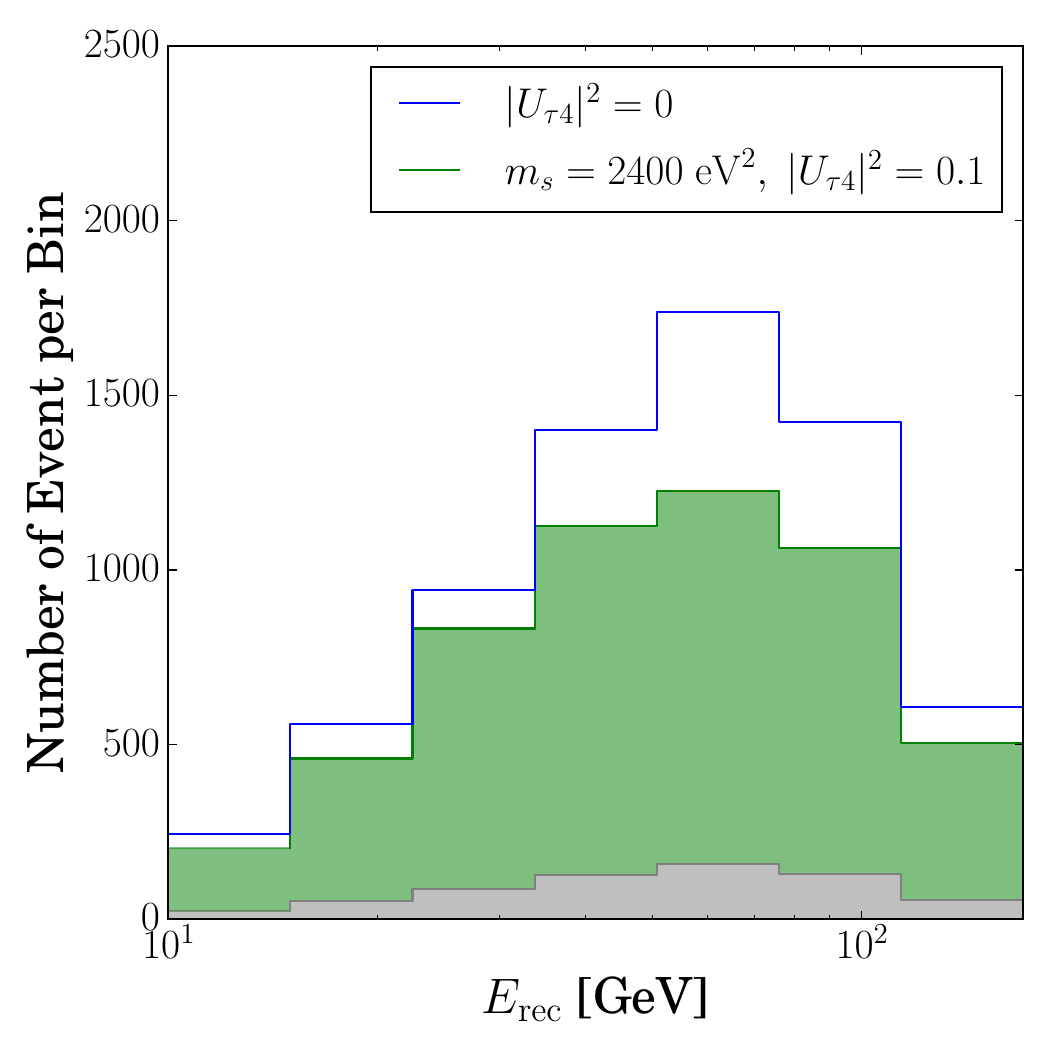}
    \end{center}
    \caption{Number of expected event per bin with the mixing $|U_{\tau 4}|^2=0.1$, assuming  10\% background (grey color with $R_\mathrm{s/b}=10$) for  $\Delta m_{41}^2=600$ (left), and $2400\, \ev^2$ (right). For a comparison, blue plot shows the most-probable pseudo data  for 3 flavor model without sterile neutrino mixing.}
    \label{fig:binned}
\end{figure}

To consider the background, we use the tau neutrino spectrum divided by signal-to-background ratio $R_\mathrm{s/b}$ in 3 flavor model~\footnote{For an appropriate treatment of the background, full MC simulation of the experiment and statistical analysis of MC data are needed, which is beyond our scope.}. Similar with $A$ and $\alpha_i$, we included a overall uncertainty $B\in \boldsymbol\lambda$ and shape uncertainties $ \beta_i\in \boldsymbol\lambda$ in the background. Therefore, the expected number of the background at $i$-bin is written as
\begin{equation}
    b_{i} = R_\mathrm{s/b}^{-1}(1+B)(1 + \beta_i) \int_{E_i^\text{low}}^{E_i^\text{high}}\left.\frac{dN}{dE_\mathrm{rec}}\right|_{|U_{\tau 4}|^2=0}dE_\text{rec}.
\end{equation}
Here, the background is independent of sterile neutrino parameters, and $R_\mathrm{s/b}$ is chosen as $10$ or $1$ in later figures.

A logarithmic scale is used for our bins, beginning at $10\ \mathrm{GeV}$ and each bin ends at 1.5 times its starting energy, with a total of 7 bins. In Fig.~\ref{fig:binned}, we show the number of expected events per bin (green) after 5 years operation with the mixing $|U_{\tau 4}|^2=0.1$, assuming 10\% background ($R_\mathrm{s/b}=10$ with grey color) for  $\Delta m_{41}^2=600\ \mathrm{eV}^2$ (left) and $2400\ \ev^2$ (right). For a comparison, the blue plot shows the most probable pseudo-data for the 3-neutrino model.

\begin{figure}
    \centering
    \begin{center}
        \includegraphics[width = .49\textwidth]{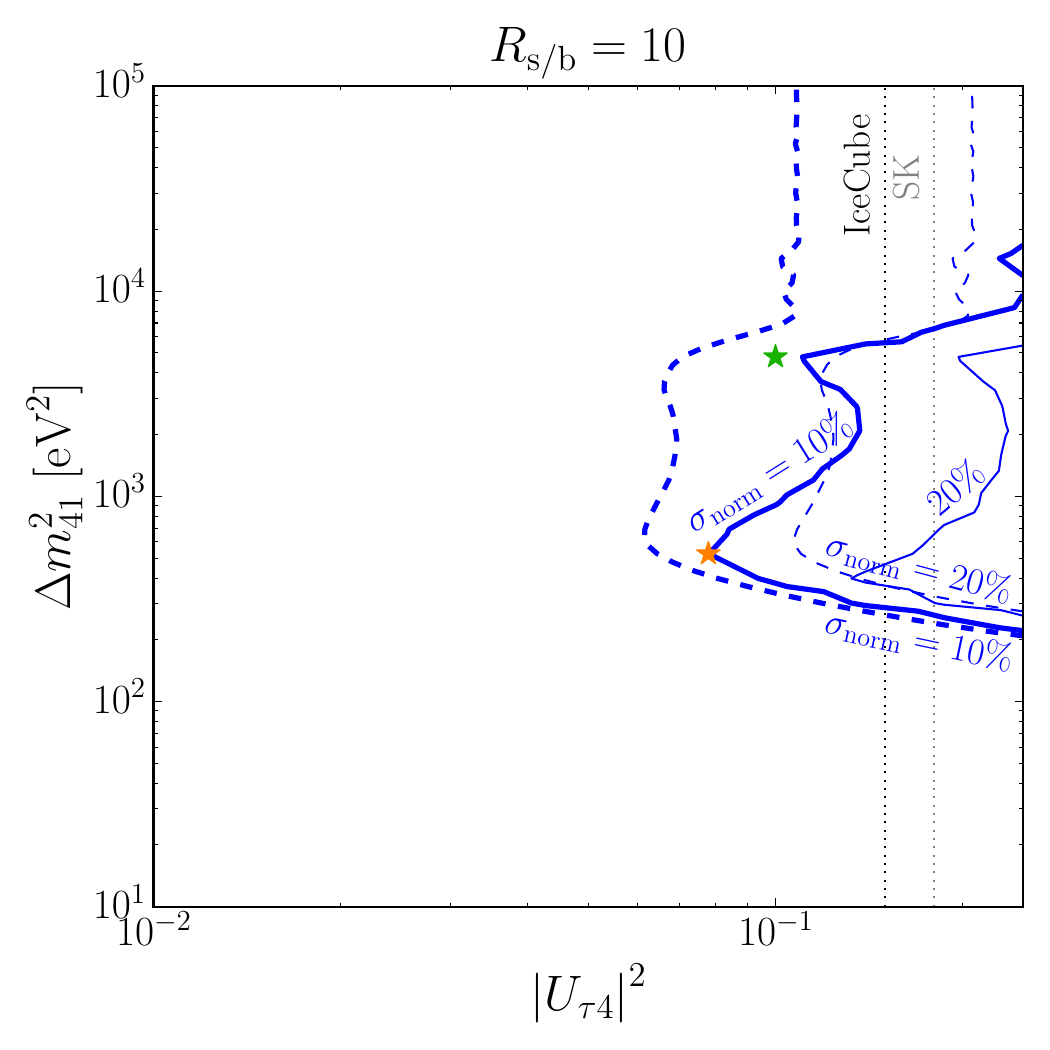}
        \includegraphics[width = .49\textwidth]{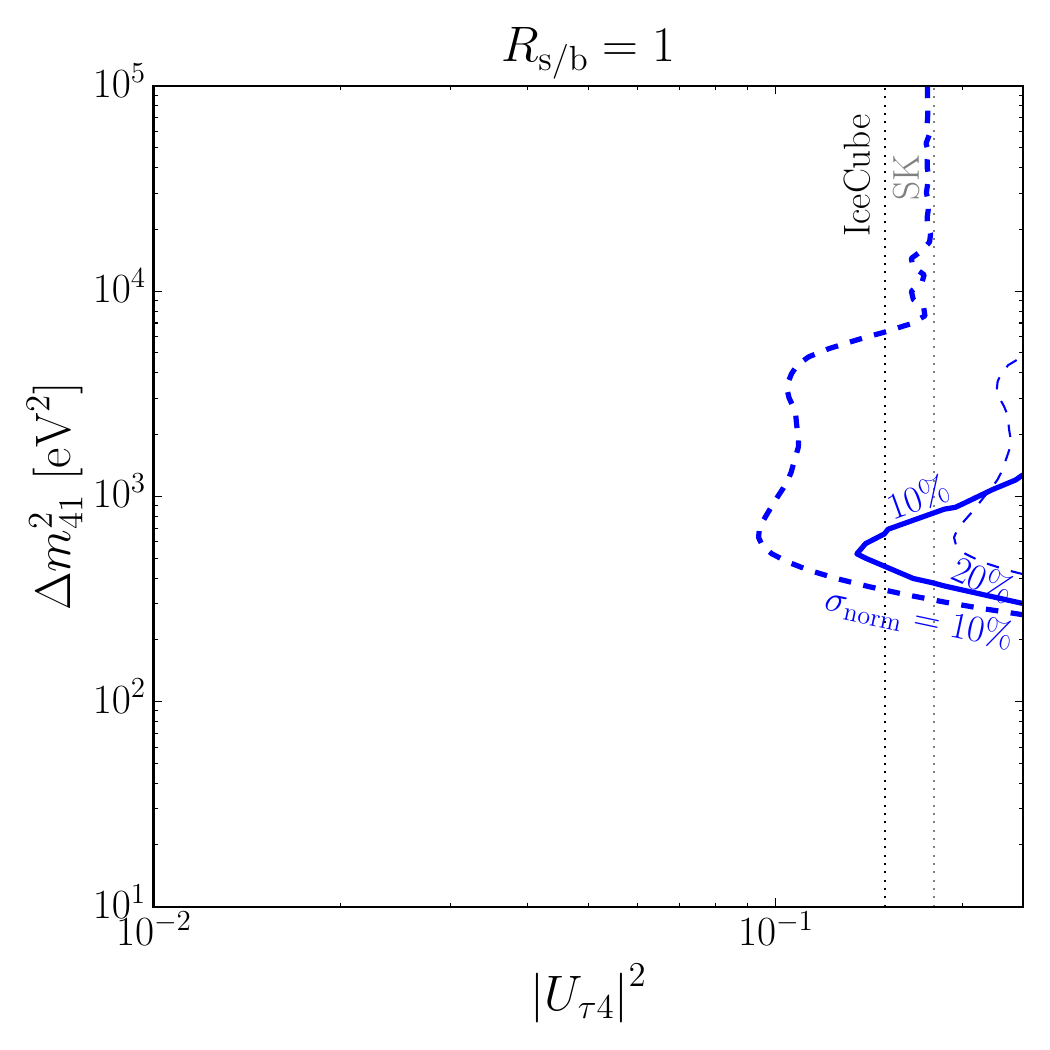}
    \end{center}
    \caption{The expected sensitivity after 5 year observation at SHiP on the plane of $(|U_{\tau 4}|^2,\Delta m^2_{41})$ under the assumption of $\theta_{14}=\theta_{24}=0$, for different number of background $R_\mathrm{s/b}=10$ (left) and $1$ (right). Blue solid (dashed) lines are the constraint of $90\%$ CL using the profiled FC method (Wilks' theorem), with uncertainty $\sigma_\mathrm{norm}= 10\%$ (thick lines) or $20\%$ (thin lines). The vertical dashed lines show the existing constraints from IceCube-DeepCore~\cite{IceCube:2017ivd} (black) and Super-Kamiokande~\cite{Super-Kamiokande:2014ndf} (grey), respectively. To check the validity of Wilks' theorem, CDFs of $\Delta\chi^2$ at orange and green stars on the left figure are drawn in Fig.~\ref{fig:CDF_NSND}.}
    \label{fig:sensitivity_NSND}
\end{figure}

To find a CL, as a first method, we use Wilks' theorem~\cite{10.1214/aoms/1177732360}, which points out that under certain conditions, the probability distribution of $\Delta\chi^2$ follows a $\chi^2$ distribution with the same number of degrees of freedom as $\boldsymbol\theta$, which is 2 in this study since  $\boldsymbol\theta= (\Delta m_{41}^2,|U_{\tau4}|^2)$. However, in neutrino oscillation models, the probability distribution of $\Delta\chi^2$ does not necessarily follows a $\chi^2$ distribution. Therefore, as a second method, we utilized the profiled FC method~\cite{NOvA:2022wnj} which is a variant of FC method~\cite{Feldman:1997qc} for a case with nuisance parameters. After choosing $\boldsymbol O$ as most-probable data in 3 flavor model, we generated $10^4$ number of pseudo-data by MC simulation on each point $\boldsymbol\theta$ on the map of model parameters, assuming $\boldsymbol\lambda = \mathrm{argmax}_{\boldsymbol\lambda} L(\boldsymbol\theta,\boldsymbol\lambda|\boldsymbol O)$. By calculating $\Delta\chi^2$ of each set of pseudo-data, we find its probability distribution which is denoted as $f(x)$.

From the probability distribution of $\Delta\chi^2$, the CL of the data $\boldsymbol O$ is defined as follows:
\begin{equation}\label{eq:defCL}
    \mathrm{CL} = \int_0^{\Delta\chi^2(\boldsymbol O)}f(x) dx,
\end{equation}
where $\Delta\chi^2(\boldsymbol O)$ is $\Delta\chi^2$ value of the data $\boldsymbol O$. Here, the definition of a CL indicates the possibility that other possible outcome from the same experiment would give a better fit, more than the data $\boldsymbol O$. The range of integration in Eq.~(\ref{eq:defCL}) starts at $x = 0$, as $\Delta\chi^2$ must be non-negative according to the definition.

\begin{figure}
    \centering
    \includegraphics[width = .49\textwidth]{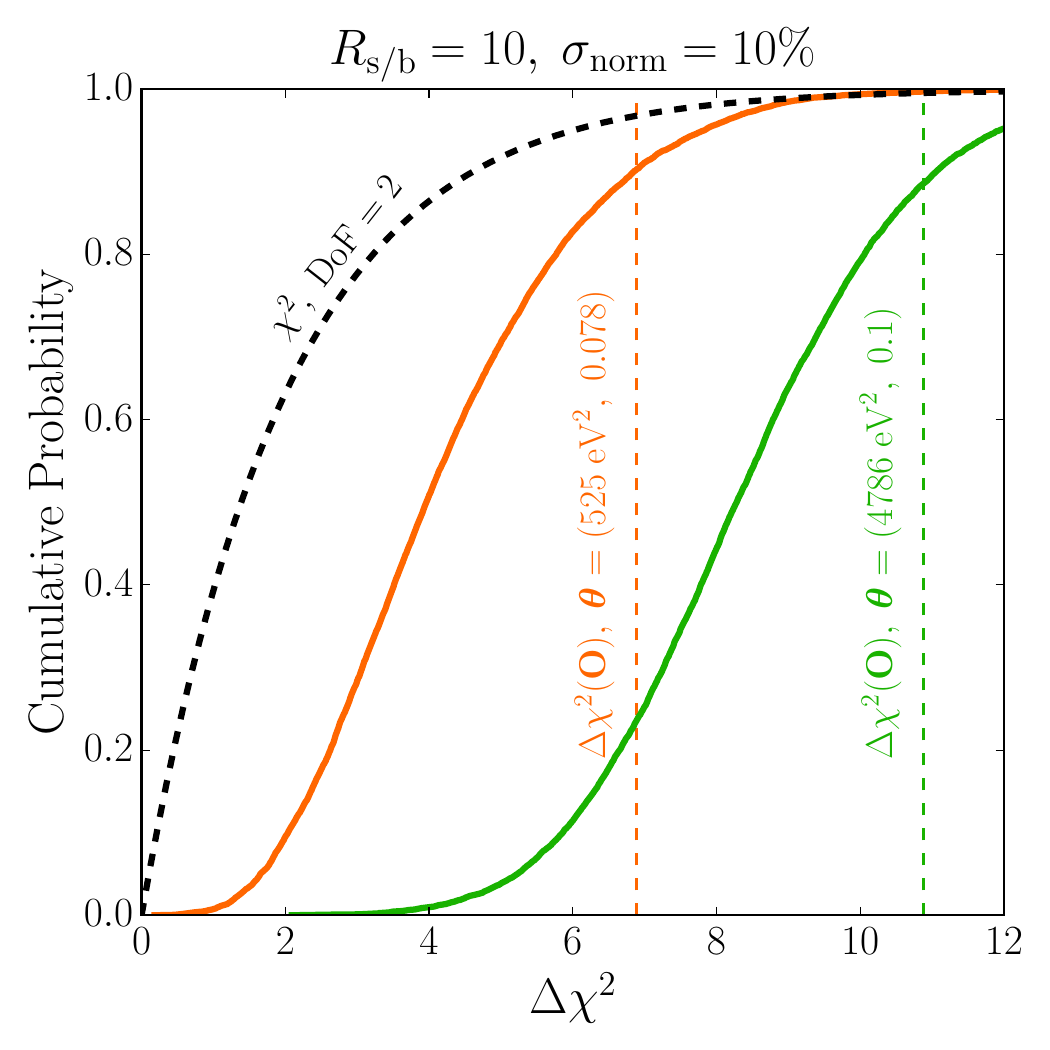}
    \caption{Cumulative distribution functions of $\chi^2$ distribution with two degrees of freedom (dashed black) and distributions of $\Delta\chi^2$ using the profiled FC method, corresponding to star shaped points in  Fig.~\ref{fig:sensitivity_NSND} with $(\Delta m_{41}^2,\ |U_{\tau 4}|^2)=(525\ \mathrm{eV}^2,\ 0.078)$ (orange), and $(4786\ \mathrm{eV}^2,\ 0.1)$ (green), for $R_\mathrm{s/b}=10$ and $\sigma_\mathrm{norm} = 10\%$. Vertical dashed lines indicate the value of $\Delta\chi^2(\boldsymbol O)$ for each point.}
    \label{fig:CDF_NSND}
\end{figure}

In Fig.~\ref{fig:sensitivity_NSND}, we show the expected sensitivity after 5 year observation at SHiP on the plane of $(|U_{\tau 4}|^2,\Delta m^2_{41})$ under the assumption of $\theta_{14}=\theta_{24}=0$, for different number of background $R_\mathrm{s/b}=10$ (left) and $1$ (right). Uncertainties of nuisance parameters are chosen as $\sigma_\mathrm{norm}= 10\%$ (thick lines) and $20\%$ (thin lines), while solid (dashed) lines are the constraint of $90\%$ CL using the profiled FC method (Wilks' theorem). Wilks' theorem is applied by drawing a contour of $\Delta\chi^2=4.61$, which corresponds to first 90\% cut of $\chi^2$ distribution with two degrees of freedom, while profiled FC method is applied through drawing a contour of $90\%$ CL using Eq.~(\ref{eq:defCL}). 
The vertical dashed lines show the existing constraints from IceCube-DeepCore~\cite{IceCube:2017ivd} and Super-Kamiokande~\cite{Super-Kamiokande:2014ndf}, respectively. 

From the left window in Fig.~\ref{fig:sensitivity_NSND}, we can find the sensitivity using profiled FC method could be $|U_{\tau 4}|^2\sim 0.08\, (0.1)$ for $\Delta m^2_{41} \sim 500 \ev^2$ with uncertainty 10\%\,(20\%), respectively, after 5 year operation of SHiP experiment with 10\% background. For $\Delta m^2_{41}\gtrsim 500 \ev^2$, in general, there is inconsistency between the sensitivities from Wilks' theorem and the profiled FC method, indicating that the probability distribution function $f(x)$ on Eq.~(\ref{eq:defCL}) from the profiled FC method does not follow $\chi^2$ distribution. 
%If $\sigma_\mathrm{norm} = 20\%$ and $R_\mathrm{s/n} =1$, the sensitivity from the profiled FC method gets weaker than the constraints from Super-Kamiokande and IceCube.

In Fig.~\ref{fig:CDF_NSND}, we show Cumulative Distribution Function (CDF)s of $\Delta\chi^2$ using profiled FC method corresponding to two star-shaped points in Fig.~\ref{fig:sensitivity_NSND} (solid orange and green), with that of $\chi^2$ distribution with two degrees of freedom (dashed black). The value of $\Delta\chi^2(\boldsymbol O)$ is displayed with vertical  dashed lines with the corresponding colors. We can check that for these points the Wilks' theorem is not applied well and thus the sensitivities between profile FC and Wilks' theorem show discrepancy.

%To check the validity of Wilks' theorem, Cumulative Distribution Function (CDF)s of $\Delta\chi^2$ at orange and green stars on the left window in Fig.~\ref{fig:sensitivity_NSND} are drawn in Fig.~\ref{fig:CDF_NSND} with the same color as stars in Fig.~\ref{fig:sensitivity_NSND}. The value of $\Delta\chi^2(\boldsymbol O)$ is displayed as dashed lines with the same colors, while to compare with Wilks' theorem, $\chi^2$ distribution with two degrees of freedom is drawn as a dotted black line. Gaps between CDFs from the profiled FC method and $\chi^2$ distribution show that Wilks' theorem in this case gives bad performance, as we seen in Fig.~\ref{fig:sensitivity_NSND}.

\begin{figure}
    \centering
    \begin{center}
        \includegraphics[height = .25\textwidth]{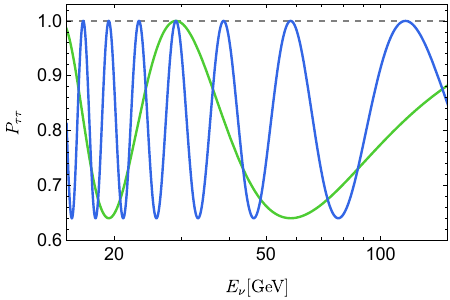}
        \includegraphics[height = .25\textwidth]{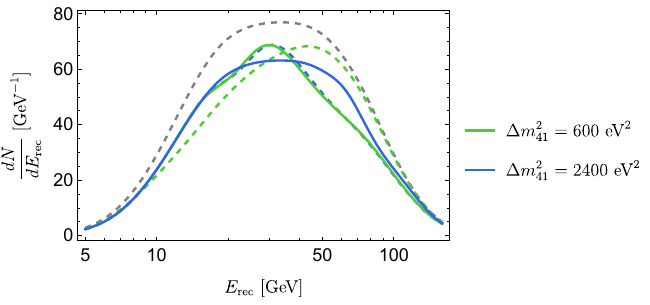}
    \end{center}
    \caption{(Left) Survival probability of tau neutrinos as a function of $E_\nu$ at $120\ \mathrm{m}$ baseline. (Right) Differential number of $\nu_\tau$ CC events at FSND , assuming the same number of event as expected in NSND, \textit{i.e.} $R_\mathrm{F/N}=100\%$. In both figures, the green line stands for $\Delta m^2_{41}=600\ev^2$  and blue line for $\Delta m^2_{41}=2400\ev^2$. In the right figure, the dashed lines are the spectrum at NSND for the corresponding mass difference, with black dashed line for the spectrum without oscillation with sterile neutrino.}
    \label{fig:FSND}
\end{figure}

%%%%%%%%%%%%%%%%%%%%%%%%%%%%%%%%%%%%%
\section{Adding a Far Detector: NSND + FSND }
\label{sec:FSND}
%%%%%%%%%%%%%%%%%%%%%%%%%%%%%%%%%%%%%

In this section, we consider an additional SND with longer baseline, which we call far SND (FSND), to enhance the sensitivity.
The distance from the proton target to FSND we assumed is $120\ \mathrm{m}$, therefore FSND will be placed after Hidden Sector Decay Spectrometer in the current design of SHiP experiment. For comparison, we refer to the SND positioned closer to the proton target as near SND (NSND). Since the disappearance pattern of tau neutrinos from `3+1' model is dependent on the baseline while nuisance parameters are not, introducing FSND gives more control on nuisance parameters.

For simplicity, we assume that the projected geometry of FSND is the same as NSND, but the vertical length of FSND can be adjusted. Depending on the vertical length of FSND, we consider several different acceptance of FSND, quantified by a parameter $R_\mathrm{F/N}$ which denotes the ratio of $\nu_\tau$ CC events at FSND compared to NSND. In Fig.~\ref{fig:FSND}, we show the survival probability of tau neutrinos at the $120\ \mathrm{m}$ baseline (left) and differential number of $\nu_\tau$ CC events at FSND (right) with $\Delta m^2_{41}=600\ev^2$ (green) and $\Delta m^2_{41}=2400\ev^2$ (blue) if FSND has the same expected number of events as NSND, \textit{i.e.} $R_\mathrm{F/N}=100\%$. For comparison, the differential number of $\nu_\tau$ CC events at NSND from Fig.~\ref{fig:fluxNSND} are shown with dashed lines with corresponding colors. We can see the clear change of spectrum from NSND to FSND.

%\cred{$\beta_i$ with the same value at NSND and FSND. if they are used differently?}

\begin{figure}
    \centering
    \begin{center}
        \includegraphics[width = .49\textwidth]{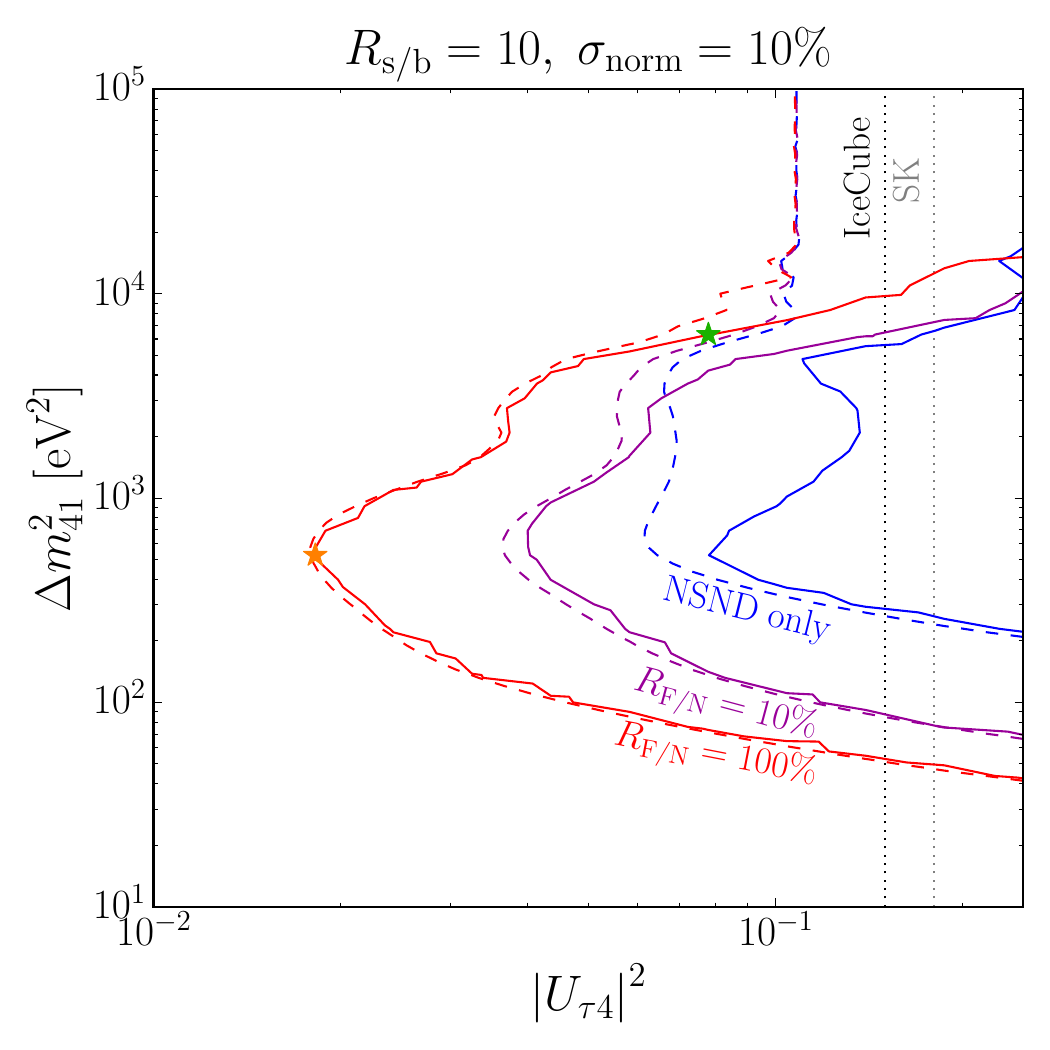}
        \includegraphics[width = .49\textwidth]{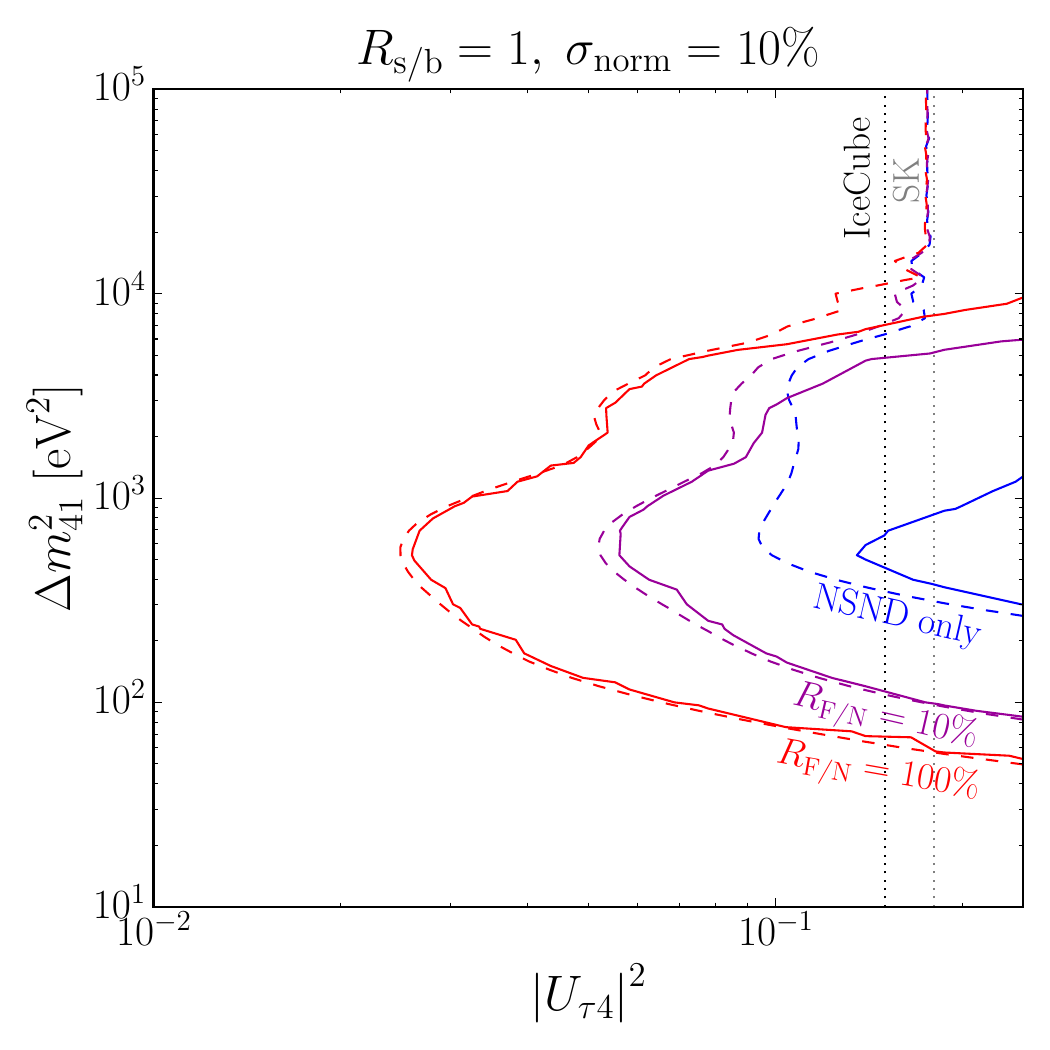}
    \end{center}
    \begin{center}
        \includegraphics[width = .49\textwidth]{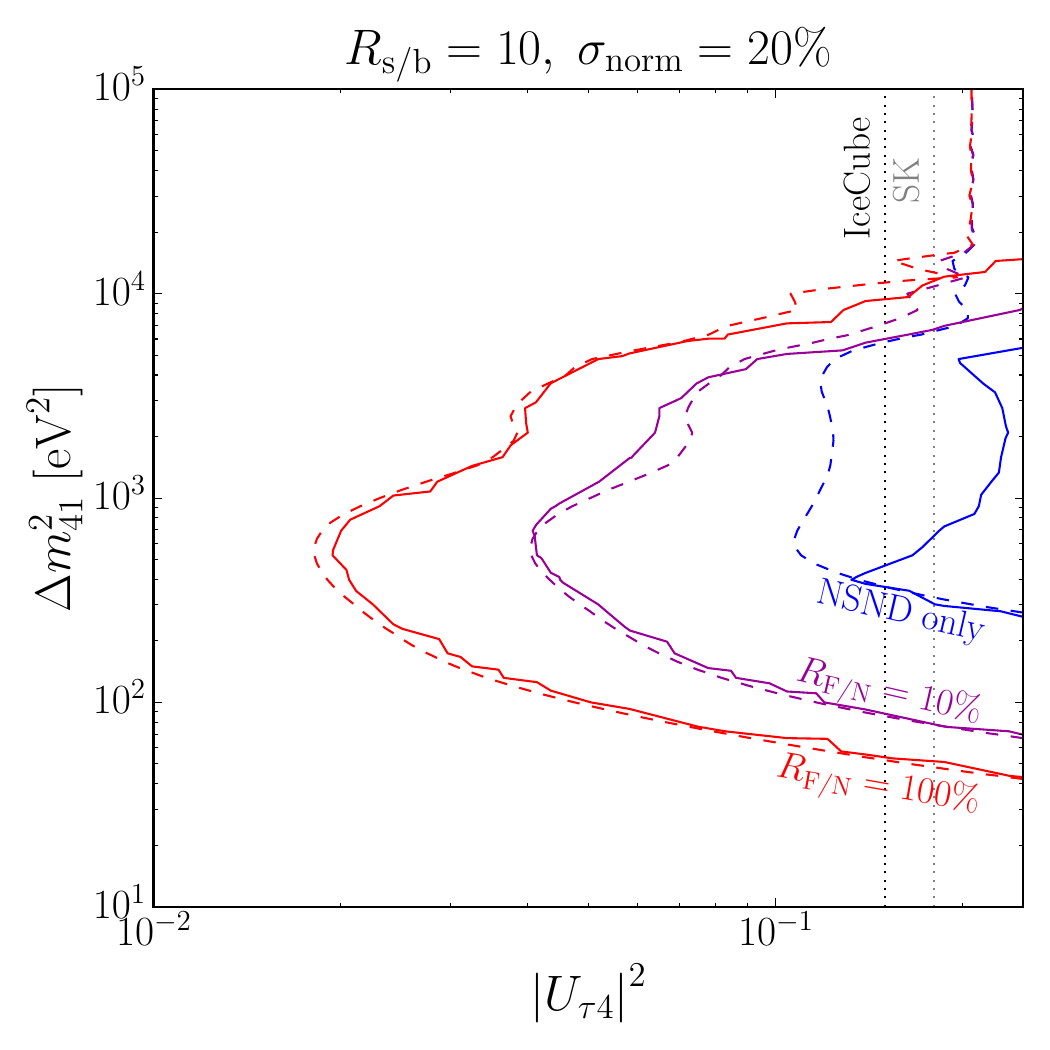}
        \includegraphics[width = .49\textwidth]{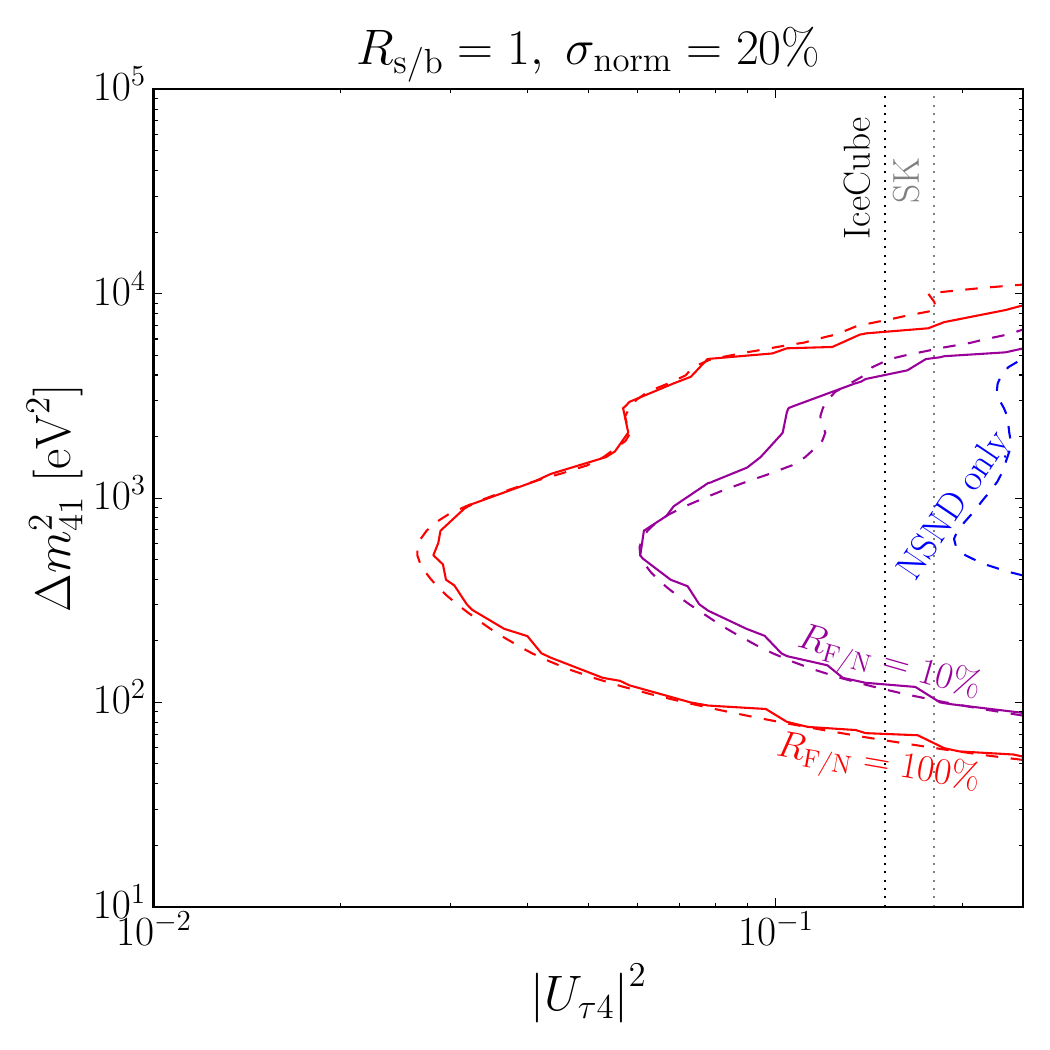}
    \end{center}
    \caption{The same as Fig.~\ref{fig:sensitivity_NSND} but with the combined analysis with both NSND and FSND for different $R_\mathrm{F/N}$ values, 10\% (purple) and 100\% (red) as well as NSND only (blue).
    In the upper (lower) windows 
    $\sigma_\mathrm{norm}=10\%$ (20\%),
    and in the left (right) windows $R_{s/b}=10$ ($1$) is used. 
    Like Fig.~\ref{fig:sensitivity_NSND}, solid (dashed) lines indicates that the profiled FC method (Wilks' theorem) is used to find the CL.}
    \label{fig:sensitivity_FSND}
\end{figure}

In Fig.~\ref{fig:sensitivity_FSND}, we show the expected sensitivity to $|U_{\tau 4}|^2$ and $\Delta m^2_{41}$ with the combined data from NSND and FSND for different geometric acceptance at FSND,  $R_\mathrm{F/N} = 10\%$,  $100\%$, and also include results from NSND only case for comparison. In the upper (lower) windows $\sigma_\mathrm{norm}=10\%$ (20\%) is used, and in the left (right) windows $R_{s/b}=10$ ($1$) is used. 
Compared to the NSND only case, both methods gives similar sensitivity near $\Delta m^2_{41} \sim 10^3\ \mathrm{eV}^2$,   
%the sensitivity with non-zero $R_\mathrm{F/N}$ gives a robust sensitivity to `3+1' model since both Wilks' theorem and the profiled FC method give similar sensitivities near $\Delta m^2_{41} \sim 10^3\ \mathrm{eV}^2$. Sensitivities from the profiled FC method
and it merely depends on $\sigma_\mathrm{norm}$, meaning that the uncertainty from nuisance parameters can be relaxed. If $R_\mathrm{F/N} = 100\%$ and $R_\mathrm{s/b} = 10\,(1)$, the combined sensitivity from NSND+FSND can be  $|U_{\tau4}|^2 \sim 0.02\,(0.03)$ near $\Delta m^2_{41}\sim 500\ \mathrm{eV}^2$.

\begin{figure}
    \centering
    \includegraphics[width = .49\textwidth]{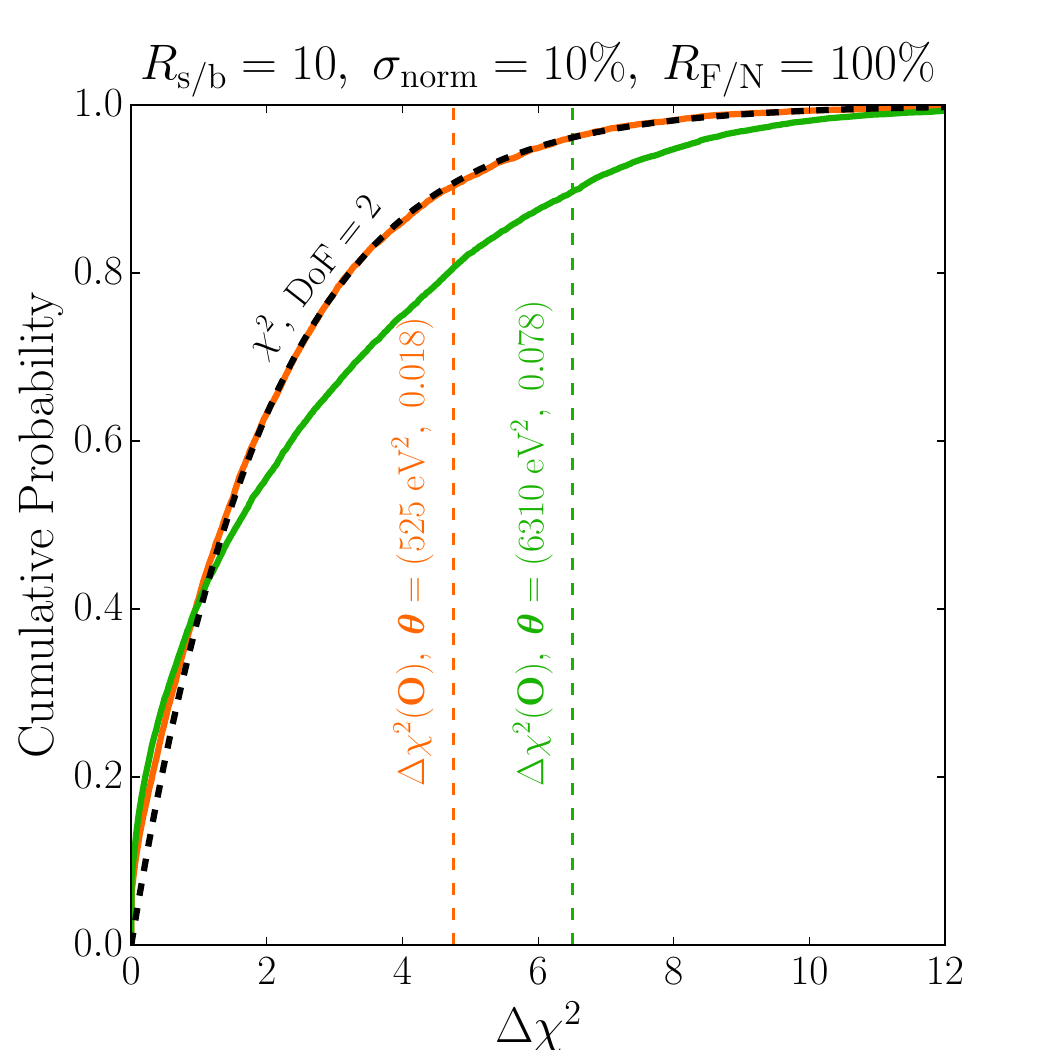}
    \caption{CDFs of $\chi^2$ distribution with two degrees of freedom (dashed black) and distributions of $\Delta\chi^2$ using the profiled FC method, corresponding to star shaped points in  Fig.~\ref{fig:sensitivity_FSND} with  $(\Delta m_{41}^2,\ |U_{\tau 4}|^2) = (525\ \mathrm{eV}^2,\ 0.018)$ (orange), and $(6310\ \mathrm{eV}^2,\ 0.078)$ (green), for $R_\mathrm{s/b}=10$, $\sigma_\mathrm{norm} = 10\%$ and $R_\mathrm{F/N}=100\%$. Vertical dashed lines indicate the value of $\Delta\chi^2(\boldsymbol O)$ for each point.}
    \label{fig:CDF_FSND}
\end{figure}

In Fig.~\ref{fig:CDF_FSND}, we show the CDFs of $\Delta\chi^2$ at orange and green stars on the upper-left window in Fig.~\ref{fig:sensitivity_FSND}. As can be seen from this figure, the CDFs from both method
are consistent very well for the orange point, which explains the similar sensitivity with both methods in Fig.~\ref{fig:sensitivity_FSND} for $\Delta m_{41}^2\lesssim 5000 \ev^2$.
For larger $\Delta m_{41}^2$, sensitivities from both methods start to mismatch, as can be seen with the CDFs for the point of the green star.

%%%%%%%%%%%%%%%%%%%%%%%%%%%%%%%
\section{Conclusion}
\label{con}
%%%%%%%%%%%%%%%%%%%%%%%%%%%%%%%
In this paper, we considered tau neutrino disappearance for the first time in the high energy fixed target experiment of SHiP, where several thousand tau neutrinos are expected to be observed after 5 years operation. This  gives good environment for the tau neutrino oscillation experiment. 

We studied sterile neutrino oscillation in `3+1' model and derived possible sensitivity to the mixing and mass difference using the energy spectrum of the tau (anti)neutrinos. With NSND only, the sensitivity would be $|U_{\tau 4}|^2\sim 0.08\, (0.1)$ near $\Delta m^2_{41} \sim 500 \ev^2$, for $\sigma_\mathrm{norm} = 10\%\,(20\%)$ and $R_\mathrm{s/b} = 10$ using profiled FC method. We find that this is slightly weaker than the sensitivity assuming the Wilks' theorem is applied.

%. Sensitivities using the profiled FC method, on the other hand, is weaker than Wilks' theorem
%, and we due to the CDFs on two sample points. 

%For the case with $\sigma_\mathrm{norm} = 20\%$ and $R_\mathrm{s/b} = 1$, the expected sensitivity of NSND is weaker than other constraints from atmospheric neutrino experiments.

To enhance the sensitivity, we suggested additional FSND which can be put after Hidden Sector Decay Spectrometer. It can play a role of far detector and its spectrum can be directly compared to that of NSND. From Fig.~\ref{fig:sensitivity_FSND}, NSND+FSND gives a robust sensitivity that does not depend on the choice of statistical method. Also, the constraints barely change with the choice of  $\sigma_\mathrm{norm}$. If $R_\mathrm{F/N} = 100\%$, the combined data gives the sensitivity $|U_{\tau 4}|^2 \sim 0.02\, (0.03)$ near $\Delta m_{41}^2\sim 500 \ev^2$ with $R_{s/b}=10\, (1)$ .

%\cblue{possibility to cancel out the systematic uncertainty between NSND and FSND.}

%understanding of the systematics is crucial to improve the sensitivity.

\acknowledgments
The authors deeply appeieciate Yu Seon Jeong for useful discussion. 
The authors were supported by the National Research Foundation of Korea (NRF) grant funded by the Korea government (MEST) 
2021R1A2C2011003, 2021R1F1A1061717, and 2022R1A2C100505.

%\paragraph{Note added.} This is also a good position for notes added after the paper has been written.

% Bibliography

%% [A] Recommended: using JHEP.bst file
\bibliographystyle{JHEP}
\bibliography{biblio.bib}

\end{document}